\documentclass[12pt]{article}
 \usepackage{graphicx,epsfig}
 \usepackage{amssymb}

\begin{document}
 \begin{center}

 {\Large Recoil Products from $p+^{118}{\rm Sn}$ and $d+^{118}{\rm Sn}$ at
3.65 GeV/A
}

\vspace{8mm}
A.R. Balabekyan \footnote{ Yerevan State University, Armenia}$^{,*}$,
A.S. Danagulyan $^{1}$,
J.R. Drnoyan $^1$,
G.H. Hovhannisyan $^1$,\\
N.A. Demekhina \footnote{ Yerevan Physics Institute, Armenia},\\
J. Adam \footnote{JINR, Dubna, Russia}$^{,}$\footnote{ INF AS R\v{e}z, 
         Czech Republic}, 
V.G. Kalinnikov $^3$, 
M.I. Krivopustov $^3$,
V.S. Pronskikh $^3$, 
V.I. Stegailov $^3$,\\
A.A. Solnyshkin $^3$, 
P. Chaloun $^{3,4}$, 
V.M. Tsoupko-Sitnikov $^3$\\
S.G. Mashnik \footnote{ Los Alamos National Laboratory,
          Los Alamos,USA},
K.K. Gudima \footnote{ Institute of Applied Physics, Academy of 
     Science of Moldova, Chi\c{s}in\u{a}u} \\

\vspace{2mm}

\end{center}
 \begin{center}
 {\bf Abstract}
 \end{center}
The recoil properties of the product nuclei from the interaction
of 3.65 GeV/nucleon protons and deuterons from 
the Nuclotron and Synchrophasotron of
the Laboratory of
High Energies (LHE), Joint Institute for Nuclear Research (JINR)
at Dubna
with a $^{118}$Sn target have been  studied using catcher foils.
The experimental data were analyzed using the mathematical
formalism of the standard two-step vector model. The 
results for protons are compared with those for deuterons. 
Our experimental results were compared to three
different Los Alamos versions of the Quark-Gluon String Model (LAQGSM).
The forward velocity v and the
recoil nuclei kinetic energy increases linearly with increasing
mass loss of the target $\Delta A$, but seems to change its slope
at around $\Delta A=60$. It seems that light- and medium-mass
products are produced partly by a fragmentation mechanism.

\section{Introduction}

The investigation of interactions of high-energy particles with
complex nuclei by the induced activity method is limited to 
measurements on the large residual nuclei that remain at the end
of reactions. The study of large residual nuclei usually
involves measurements of either their excitation functions or
their recoil properties.

In order to determine the recoil properties of nuclei 
``thick-target thick-catcher" experimens are used. In such
experiments, the thicknesses of the target and catcher foils are
larger than the longest recoil range. The quantities measured are
the fractions $F$ and $B$ of product nuclei that recoil out of the
target foil into the forward and backward directions, respectively.

The results of the experiment are usually proceeded by the
standard two-step vector representation \cite{sugar}--\cite{win2}.
The following assumptions are made in this model: 

\hspace*{10mm}(1) In the first step, the incident particle interacts 
with the target nucleus to form an excited nucleus with velocity
$v$, momentum $P$, and excitation energy $E^*$. 

\hspace*{10mm}(2) In the second step, the excited nucleus loses
mass and excitation energy to form the final recoiling nucleus
with an additional velocity $V$, which in general will have a
distribution of values and directions. 

Usually, additional assumptions are made in most experiments:

\hspace*{10mm}(a) The quantities $v$ and $P$ in the first step are
constant \cite{kort} and lie in the forward direction. 

\hspace*{10mm}(b) The velocity in the second step is isotropic.

The results of the recoil experiments depend on the range-energy
relation of the recoiling nuclei. It is convenient to express this
relation as  \cite{win1}:
\begin{equation}
R=kV^n ,
\end{equation}
where $R$ is the mean range (corresponding to $V$) of the recoil in
the target material, $k$ and $n$ are constants and can be
evaluated from tables of ranges of nuclei recoiling into
various materials \cite{data}. The following relations are used
for the forward and backward fractions:
\begin{equation}
FW=\frac{1}{4}R [1+\frac{2}{3} (n+2) \eta + \frac{1}{4} (n+1)^2
\eta^2];  \,\,\,\,\,
 BW=\frac{1}{4} R[1-\frac{2}{3} (n+2) \eta +
\frac{1}{4} (n+1)^2 \eta^2]
\end{equation}
where $\eta = v/V $  and $W$ is the target thickness in $mg/cm^2$.

The reaction product mean ranges ($R$) and the velocities
transferred to residuals on the first ($v$) and second ($V$) steps of
reaction are calculated using the following
expressions \cite{win1}:
\begin{equation}
 F/B=\frac{1+\frac{2}{3} (n+2) \eta + \frac{1}{4} (n+1)^2
\eta^2}{1-\frac{2}{3} (n+2) \eta + \frac{1}{4} (n+1)^2 \eta^2};
\,\,\,\,\,\,
 R=2W(F+B)/(1+\frac{1}{4}(n+1)^2 \eta^2)
\end{equation}

The recoil properties of non-fission reactions induced by protons
with energy 1 GeV and above were analyzed with the two-vector model in
Ref.\ \cite{win3}. Systematic deviations of the resulting
parameters were attributed to the presence of fragmentation. The
kinematic properties of radionuclides formed in photospallation
reactions on complex nuclei at intermediate energy and a comparison
with proton-nuclear reactions were made in Ref.\ \cite{haba}. 
We studied the recoil properties of nuclei produced in the
photospallation of $^{65}$Cu in Ref.\ \cite{arak}. The purpose of 
the present work is to investigate the
kinematic properties of product nuclei formed in the target
$^{118}$Sn bombarded with 3.65 GeV/nucleon protons
and deuterons using catcher foils.

\section{Experimental setup and results}

Targets of enriched tin isotope $^{118}$Sn were irradiated at the
Nuclotron and Synchrophasotron of the 
LHE,
JINR
by proton
and deuteron beams with energies of 3.65 GeV/nucleon. Irradiations
were of 6.42 hour for the proton beam and 1.083 hour for the
deuteron beam. The deuteron beam had an elliptic form with axes of
3 and 2 cm. The proton beam had round form with a diameter of 2 cm.
For beam monitoring, we employ the reactions
$^{27}$Al$(d,3p2n)^{24}$Na and $^{27}$Al$(p,3pn)^{24}$Na whose
cross sections are taken as of $14.2\pm0.2$mb \cite{8} and
$10.6\pm0.8$mb \cite{9}, respectively. From the monitoring
reactions the following beam intensities were obtained:
$0.768\times10^{13}$ d/hour and $0.114\times10^{13}$ p/hour. The
total beam fluences were $3.21\times10^{13}$ protons and
$2\times10^{13}$ deuterons.

The target consisted of a high-purity target metal foil of size
20x20 mm$^2$ sandwiched exactly by one pair of Mylar foils
of the same size, which collected the recoil nuclei in the forward
or backward directions with respect to the beam. The enrichment of
the target was 98.7 \%, the thickness of each target foil was 66.7
$mg/cm^2$ and the number of target piles was 15. The whole stack,
together with an Al beam-monitor foil of 140 mg/cm$^2$ thickness was
mounted on a target holder and irradiated in air.

After irradiation the target foils and all of the forward and
backward catcher foils from one target pile were collected
separately, and assayed for radioactivities nondestructively with
high-purity Ge detectors at LNP, JINR for one year. The radioactive
nuclei were identified by characteristic $\gamma$ lines and by
their half-lives. The spectra were evaluated with the code package 
DEIMOS32 \cite{frana}.

The kinematic characteristics of thirty product nuclei were
obtained for deuteron- and proton-induced reactions. The relative
quantities of the  forward- and backward-emitted nuclei  (relative 
to the beam direction) were calculated from relations:
\begin{equation}
F=N_F/(N_t+N_F+N_B); \,\,\,\,  B=N_B/(N_t+N_F+N_B)
\end{equation}
where $N_F$ , $N_B$ , $N_t$ are the numbers of nuclei emitted in
forward and backward catchers and formed in target foils,
respectively. The recoil parameters obtained in these experiments
are the forward-to-backward ratio, $F/B$, and the mean range,
$2W(F+B)$. [The mean range of the recoils is somewhat smaller than
$2W(F+B)$, but it is conventional to refer to the latter quantity
as range]. The mathematical formalism of the standard two-step
vector model \cite{win1} was used to proceed the
experimental results. The parameters  k and n in equation (1) are
obtained by fitting the range dependence on energy of accelerated
ions within the region from 0.025 to 5 MeV/nucleon \cite{range}.
It is possible to calculate $\eta$ and $v$ from the equation
(3), knowing the $F/B$ ratio of the experiment.

Our experimental results are shown in
Tables 1 and 2. We note that uncertainties concerning definite 
quantities in our tables are not listed to keep the tables concise. 
These uncertainties are about 10--15\%. As
is seen from the tables, the kinetic energies of product nuclei from
proton-induced reactions are higher than are the ones induced by 
deuterons.  Probably, protons are more effective agents
of linear momentum transfer on a per-nucleon basis when compared
with deuterons \cite{laurent,Bron}.

As shown in Fig.\ 1, the ratios $F/B$ for both proton- and
deuteron-induced reactions are of the order of $\sim 3 - 4$ for heavy
%
%
product nuclei and decrease to about $\sim 2$ for light residuals.
Such a dependence could be explained by different mechanisms
for the  production of nuclei in different mass regions. Light nuclei
may be produced by multifragmentation and evaporation
that lead to an isotropic distribution in the frame of excited 
residual nuclei,  while heavy nuclei are produced mainly 
via the spallation mechanism, with its products more in
the forward direction. Some of our recent studies
\cite{bal3,bal2} point to the multifragmentation mechanism
in the formation of product nuclei with light and medium mass
numbers.

Our experimental results are compared with
theoretical calculations by the LAQGSM03.01 \cite{LAQGSM0301},
LAQGSM03.S1 \cite{LAQGSM03S1}, and LAQGSM03.G1 \cite{LAQGSM03S1} models. 

LAQGSM03.01 \cite{LAQGSM0301} is the latest modification of the  
Los Alamos version of the Quark-Gluon String Model \cite{LAQGSM},
which in its turn is an improvement of the  
Quark-Gluon String Model \cite{QGSM}.
It describes reactions induced by both particles and nuclei as
a three-stage process: IntraNuclear Cascade (INC), followed
by preequilibrium emission of particles during the equilibration of the
excited residual nuclei formed during the INC, followed by
evaporation of particles from or fission of the compound nuclei.
The INC stage of reactions is described 
with a recently improved version \cite{LAQGSM0301} 
of the time-depending intranuclear cascade
model developed initially at Dubna, often 
referred in the literature simply
as the {\bf D}ubna intranuclear {\bf C}ascade {\bf M}odel, DCM
(see \cite{DCM} and references therein).
The preequilibrium part of reactions is described with
an improved version \cite{CEM03.01} of the Modified
Exciton Model (MEM) from the Cascade-Exciton Model, CEM \cite{CEM}. 
The evaporation and fission stages of reactions
are calculated with an updated and improved version of
the Generalized Evaporation Model code GEM2 by Furihata \cite{GEM2},
which considers evaporation of up to 66 types of different particles
and light fragments (up to $^{28}$Mg). If the excited residual
nucleus produced after the INC has a mass number $A \leq 11$,
LAQGSM03.01 uses a recently updated and improved
version of the Fermi Break-up model (in comparison with the
version described in \cite{QGSM}) to calculate its decay 
instead of considering a preequilibrium stage followed by 
evaporation from compound nuclei, as described above.
LAQGSM03.01 considers also coalescence of complex
particles up to $^4$He from energetic nucleons emitted during the INC,
using an updated coalescence model in comparison with the version
described in \cite{DCM}.

LAQGSM03.S1 \cite{LAQGSM03S1} is exactly the same as LAQGSM03.01,
but considers also multifragmentation of excited nuclei produced
after the preequilibrium stage of reactions, when their excitation
energy is above 2A MeV, using the Statistical Multifragmentation
Model (SMM) by Botvina et al.\ \cite{SMM} (the ``S" in the extension
of LAQGSM03.S1 stands for SMM).

LAQGSM03.G1 \cite{LAQGSM03S1} is exactly the same as LAQGSM03.01,
but uses the fission-like binary-decay model GEMINI
of Charity et al.\ \cite{GEMINI},
which considers evaporation of all possible fragments,
instead of using the GEM2 model \cite{GEM2}
(the ``G" stands for GEMINI).

As can be seen from Fig.\ 1, there is some disagreement between
experimental data and theoretical results by all three
versions of LAQGSM considered here. 
In making such a comparison, we first recognize
that the experiment and the calculations differ in that:
1)
the experimental data were extracted assuming
the ``two-step vector model" 
\cite{sugar}--\cite{win2}, 
while the LAQGSM calculations were done
without the assumptions of this model;
2) the measurements were performed on foils (thick targets),
while the calculations were done for interactions
of protons/deuterons with nuclei (thin targets).
These differences must be considered before assessing
possible deficiencies in the models.

It is interesting to note that the measured forward velocity v increases
practically linearly with the increase of 
$\Delta A$ ($\Delta A = A_{targ}-A_{res}$,
where $A_{targ}$ is the mass number of target and
$A_{res}$ is the mass number of product nuclei), but seems to change
it slope at around $\Delta A=60$. Comparison of these results
with theoretical calculations shows that this is due to different
mechanisms for the production of the measured nuclei (Fig.\ 2).

Fig.\ 3 shows the dependence of the fragment kinetic energy, $T_{kin}$, on
the fractional mass loss $\Delta A/A$. Cumming and B\"{a}chmann
\cite{cum} and Winsberg \cite{win2} have shown that $T_{kin}$ should
increase linearly with $\Delta A/A$ for reactions in which the
velocity of the product is due to the vectorial addition of the
randomly directed recoil velocities resulting from particle
emission. 
For comparison, Fig.\ 3 shows also data measured for a Ag target by
Porile et al.\ (see Table II in \cite{win3} and references therein).
One can see that the Ag results agree well with our
current $^{118}$Sn data.

The deviation from a linear trend seen in Fig.\ 3 for large fractional
mass losses probably indicates a change in mechanism
of the production of light nuclides. The formation
of light fragments from highly-excited nuclei is of a permanent
interest in the literature and usually 
considers a multi-body breakup \cite{botv1}. One possible
mechanism for such a process would be a simultaneous clustering of
nucleons into fragments near the liquid-gas critical point
\cite{SMM}. This process is essentially different from the
sequential evaporation process by which deep spallation products
are formed. It appears that the mean kinetic energies of the
products provide a qualitative method for distinguishing between
these two mechanism \cite{wang}.

As one can see from Fig.\ 3,
the results by LAQGSM03.S1, which 
considers multifragmentation \cite{SMM} of excited nuclei
when their excitation energy is above 2 MeV/nucleon, overestimate
significantly
the values of the measured mean kinetic energies of light products.
This could be an indication that multifragmentation
become important only at higher excitation energies, 4--5 MeV/nucleon
instead of 2 MeV/nucleon as considered by LAQGSM03.S1, in complete
agreement with the very recent ISiS measurements \cite{ISiS}.
      
The evaluated ``experimental" excitation energies of residual nuclei produced
after the first, cascade stage of reactions are shown in the last column
of Tables 1 and 2. 
The relation between the excitation
energy ($E^*$) and $v$ may be estimated as \cite{schd}:
\begin{eqnarray}
E^{*}=3.253*10^{-2}k^{'}A_T v [T_p/(T_p+2)]^{0.5} ,
\end{eqnarray}
where $E^*$  and the bombarding energy $T_p$ are expressed in
terms of $m_pc^2$. $A_T$ is the target mass in $amu$ and $v$ is in
units of (MeV/$amu)^{0.5}$. Usually, the constant $k'$ is taken as
$k^{'}=0.8$.

As is seen from tables, the excitation energies 
estimated according to Eq.\ (5) are higher than the
multifragmentation threshold $E_{th}=2-4$ MeV/nucleon
($E_{tot}=216-424$ MeV) \cite{botv2} for the light and medium
products. 
This could be an indication that light and medium fragments
are produced not only via the
evaporation mechanism but also via multifragmentation.

\section*{Acknowledgments}

The authors would like to express their gratitude to the operating personnel
of the JINR Nuclotron and Synchrophasotron for providing good beam
parameters and to thank Dr. A.J. Sierk of LANL 
for a most useful critical reading.

This work was supported partially by the US DOE.

\newpage
 \begin{center}
 \textbf{Table 1}. Kinematic characteristics of product nuclei from
 deuteron-induced reactions

 \vspace*{5mm}
 \begin{tabular}{|c|c|c|c|c|c|c|}  \hline
 Product     &   $F/B$ & $\eta$ &  $2W(F+B)$  & $T_{kin}$ (MeV)& $v$ (MeV/$amu)^{1/2}$&$E^*$ (MeV) \\ \hline
 $^{24}$Na   &  2.03  & 0.175  & 4.08$\pm$0.40 & 15.75$\pm$ 2.37  & 0.2255 & 619.62 \\ \hline
 $^{28}$Mg   &  1.77  & 0.142  & 4.49$\pm$0.70 & 19.70$\pm$ 4.80  & 0.1913 & 522.36 \\ \hline
 $^{42}$K    &  2.36  & 0.212  & 2.49$\pm$0.60 & 12.07$\pm$ 5.11  & 0.1861 & 496.32 \\ \hline
 $^{43}$K    &  2.73  & 0.246  & 1.93$\pm$0.35 & 7.54$\pm$ 2.42   & 0.1680 & 449.30 \\ \hline
 $^{44m}$Sc  &  2.00  & 0.172  & 2.49$\pm$0.58 & 12.21$\pm$4.63   & 0.1457 & 394.77 \\ \hline
 $^{52g}$Mn  &  1.97  & 0.168  & 0.94$\pm$0.21 & 2.86 $\pm$1.03   & 0.0678 & 172.55 \\ \hline
 $^{56}$Mn   &  2.39  & 0.215  & 1.67$\pm$0.61 & 6.91 $\pm$4.06   & 0.1212 & 329.11 \\ \hline
 $^{67}$Ga   &  2.18  & 0.193  & 1.51$\pm$0.44 & 6.89 $\pm$3.21   & 0.0996 & 270.13 \\ \hline
 $^{73}$Se   &  3.08  & 0.274  & 0.72$\pm$0.16 & 2.25$\pm$ 0.83   & 0.0769 & 210.08 \\ \hline
 $^{75}$Br   &  3.02  & 0.269  & 0.94$\pm$0.28 & 3.46$\pm$ 1.70   & 0.0927 & 252.67 \\ \hline
 $^{77}$Br   &  1.97  & 0.167  & 1.41$\pm$0.39 & 6.51$\pm$ 2.88   & 0.0785 & 212.61 \\ \hline
 $^{81}$Rb   &  4.09  & 0.338  & 1.14$\pm$0.14 & 4.85$\pm$ 0.98   & 0.1318 & 361.96 \\ \hline
 $^{71}$As   &  2.84  & 0.255  & 1.45$\pm$0.27 & 6.74$\pm$ 2.03   & 0.1261 & 343.88 \\ \hline
 $^{83}$Sr   &  3.39  & 0.296  & 1.11$\pm$0.21 & 4.73$\pm$ 1.45   & 0.1130 & 309.10 \\ \hline
 $^{85}$Y    &  3.51  & 0.304  & 0.89$\pm$0.12 & 3.35$\pm$ 0.72   & 0.0963 & 263.75 \\ \hline
 $^{86m}$Y   &  3.09  & 0.275  & 0.97$\pm$0.13 & 3.88$\pm$ 0.82   & 0.0934 & 254.93 \\ \hline
 $^{87m}$Y   &  3.44  & 0.299  & 0.97$\pm$0.17 & 3.88$\pm$ 1.12   & 0.1010 & 276.46 \\ \hline
 $^{86}$Zr   &  4.78  & 0.372  & 0.78$\pm$0.11 & 2.78$\pm$ 0.66   & 0.1061 & 292.67 \\ \hline
 $^{89}$Zr   &  2.96  & 0.265  & 0.88$\pm$0.16 & 3.33$\pm$0.97    & 0.0821 & 224.13 \\ \hline
 $^{90}$Mo   &  2.39  & 0.214  & 0.84$\pm$0.17 & 3.26$\pm$1.08    & 0.0657 & 178.49 \\ \hline
 $^{93m}$Mo   &  2.07  & 0.179  & 0.52$\pm$0.07 &1.46$\pm$ 0.33   & 0.0362 & 98.25 \\ \hline
 $^{90}$Nb   &  3.28  & 0.288  & 0.45$\pm$0.09 &2.38$\pm$ 0.48    & 0.0515 & 140.56 \\ \hline
 $^{94}$Tc   &  3.12  & 0.277  & 0.36$\pm$0.09 & 0.86$\pm$  0.34  & 0.0423 & 115.53 \\ \hline
 $^{95}$Tc   &  3.82  & 0.323  & 0.54$\pm$0.08 & 1.61$\pm$ 0.39   & 0.0669 & 183.76 \\ \hline
 $^{96}$Tc   &  2.06  & 0.179  & 0.36$\pm$0.06 & 0.85$\pm$ 0.21   & 0.0270 & 73.29 \\ \hline
 $^{97}$Ru   &  3.64  & 0.312  & 0.66$\pm$0.16 & 2.28$\pm$ 0.89   & 0.0764 & 209.48 \\ \hline
 $^{99m}$Rh  &  3.56  & 0.307  & 0.45$\pm$0.07 & 1.27$\pm$ 0.33   & 0.0555 & 152.21 \\ \hline
 $^{104}$Ag  &  4.46  & 0.357  & 0.23$\pm$0.05 & 0.45$\pm$ 0.14   & 0.0374 & 103.03 \\ \hline
 $^{109}$In  &  3.24  & 0.286  & 0.26$\pm$0.04 & 0.58$\pm$ 0.14   & 0.0332 & 90.91 \\ \hline
 $^{110}$Sn  &  2.00  & 0.180  & 0.21$\pm$0.05 & 0.41$\pm$ 0.16   & 0.0168 & 45.73 \\ \hline
 $^{111}$In  &  2.33  & 0.210  & 0.16$\pm$0.04 & 0.26$\pm$ 0.11   & 0.0161 & 43.80 \\ \hline
 $^{117m}$Sn &  2.11  & 0.180  & 0.25$\pm$0.05 & 0.56$\pm$ 0.18   & 0.0064 & 56.20 \\ \hline
\end{tabular}
 \end{center}

\newpage
 \begin{center}
 \textbf{Table 2}. Kinematic characteristics of product nuclei from
 proton-induced reactions

 \vspace*{5mm}
 \begin{tabular}{|c|c|c|c|c|c|c|}  \hline
 Product    &  $F/B$ & $\eta$& $2W(F+B)$ &  $T_{kin}$ (MeV)  & $v$ (MeV/$amu)^{1/2}$ & $E^*$ (MeV) \\ \hline
 $^{24}$Na  & 1.95  & 0.165 &  5.37$\pm$ 0.35 & 23.91$\pm$ 2.36  & 0.2628 & 583.06 \\ \hline
 $^{28}$Mg  & 2.03  & 0.175 &  4.61$\pm$ 0.30 & 20.54$\pm$ 2.11  & 0.2399 & 530.01 \\ \hline
 $^{42}$K   &  2.15 & 0.189 &  3.12$\pm$ 0.20 & 17.88$\pm$ 2.04  & 0.2023 & 435.23 \\ \hline
 $^{43}$K   &  2.19 & 0.194 &  2.37$\pm$ 0.14 & 10.83$\pm$ 1.16  & 0.1595 & 343.19 \\ \hline
 $^{44m}$Sc &  2.50 & 0.225 &  2.83$\pm$ 0.18 & 14.97$\pm$ 1.57  & 0.2111 & 463.51 \\ \hline
 $^{46}$Sc  &  1.94 & 0.165 &  2.94$\pm$ 0.20 & 15.60$\pm$ 1.68  & 0.1547 & 338.71 \\ \hline
 $^{48}$Sc  & 2.08  & 0.181 &  2.55$\pm$ 0.34 & 12.07$\pm$ 2.60  & 0.1467 & 321.13 \\ \hline
 $^{48}$V   &  2.21 & 0.196 &  2.56$\pm$ 0.19 & 13.67$\pm$ 1.67  & 0.1687 & 369.73 \\ \hline
 $^{52g}$Mn &  2.21 & 0.196 &  2.09$\pm$ 0.14 & 10.43$\pm$ 1.09  & 0.1415 & 310.03 \\ \hline
 $^{58}$Co  & 3.21  & 0.284 &  2.38$\pm$ 0.18 & 12.66$\pm$ 1.54  & 0.2122 & 467.95 \\ \hline
 $^{67}$Ga  & 2.72  & 0.245 &  1.66$\pm$ 0.11 &8.04$ \pm$ 0.84   & 0.1362 & 299.48 \\ \hline
 $^{71}$As  & 3.83  & 0.324 &  1.54$\pm$ 0.10 &7.39$\pm$ 0.79    & 0.1665 & 368.74 \\ \hline
 $^{73}$Se  &  3.96 & 0.331 & 0.97$\pm$ 0.06  & 3.62$\pm$ 0.37   & 0.1174 & 260.25 \\ \hline
 $^{75}$Se  &  5.17 & 0.389 & 0.98$\pm$ 0.02  & 3.77$\pm$ 0.15   & 0.1378 & 307.88 \\ \hline
 $^{77}$Br  &  3.23 & 0.285 &  0.94$\pm$ 0.06 & 3.37$\pm$ 0.35   & 0.0954 & 210.39 \\ \hline
 $^{81}$Rb  &  4.48 & 0.358 &  1.30$\pm$ 0.07 & 5.99$\pm$ 0.56   & 0.1547 & 344.22 \\ \hline
 $^{82}$Rb  &  3.68 & 0.314 &  0.73$\pm$ 0.05 & 2.39$\pm$ 0.25   & 0.0858 & 189.87 \\ \hline
 $^{83}$Sr  &  3.32 & 0.291 &  1.22$\pm$ 0.08 & 5.51$\pm$ 0.58   & 0.1199 & 264.95 \\ \hline
 $^{85}$Y   &  3.97 & 0.331 &  1.00$\pm$ 0.07 & 3.99$\pm$ 0.42   & 0.1143 & 253.54 \\ \hline
 $^{86m}$Y  &  4.7  & 0.369 &  1.21$\pm$ 0.08 & 5.51$\pm$ 0.58   & 0.1478 & 329.47 \\ \hline
 $^{86}$Zr  & 3.83  & 0.324 &  0.75$\pm$ 0.05 & 4.88$\pm$ 0.43   & 0.1179 & 272.26 \\ \hline
 $^{89}$Zr  & 3.82  & 0.323 & 1.04$\pm$0.07   & 4.36$\pm$ 0.46   & 0.1138 & 252.26 \\ \hline
 $^{90}$Mo  &  4.30 & 0.349 &  1.19$\pm$ 0.08 & 5.74$\pm$ 0.60   & 0.1401 & 311.45 \\ \hline
 $^{90}$Nb  &  4.03 & 0.335 &  0.70$\pm$ 0.05 & 2.37$\pm$0.25    & 0.0863 & 191.72 \\ \hline
 $^{93m}$Mo & 3.74  & 0.318 &  0.68$\pm$  0.05& 2.27$\pm$ 0.25   & 0.0792 & 175.55 \\ \hline
 $^{94}$Tc  & 3.91  & 0.328 &  0.51$\pm$0.03  & 1.49$\pm$ 0.16   & 0.0656 & 145.55 \\ \hline
 $^{95}$Tc  & 3.59  & 0.309 &  0.65$\pm$ 0.04 & 2.12$\pm$ 0.22   & 0.0740 & 163.99 \\ \hline
 $^{96}$Tc  &  2.93 & 0.262 &  0.43$\pm$ 0.04 & 1.12$\pm$ 0.15   & 0.0453 & 100.00 \\ \hline
 $^{97}$Ru  & 3.86  & 0.326 &  0.53$\pm$ 0.03 & 1.60$\pm$ 0.17   & 0.0666 & 147.87 \\ \hline
 $^{99m}$Rh &  4.23 & 0.346 & 0.43$\pm$ 0.03  & 1.18$\pm$ 0.12   & 0.0599 & 133.43 \\ \hline
 $^{104}$Ag &  4.00 & 0.333 & 0.25$\pm$ 0.02  & 0.50$\pm$ 0.05   & 0.0369 & 81.99 \\ \hline
 $^{109}$In & 3.94  & 0.330 & 0.22$\pm$ 0.01  & 0.42$\pm$ 0.04   & 0.0325 & 72.33 \\ \hline
 $^{110}$Sn &  3.81 & 0.323 &  0.18$\pm$ 0.01 & 0.33$\pm$ 0.03   & 0.0282 & 62.81 \\ \hline
 $^{111}$In &  3.27 & 0.288 & 0.12$\pm$  0.01 & 0.16$\pm$ 0.02   & 0.0177 & 39.25 \\ \hline
 $^{117m}$Sn&  3.27 & 0.288 &  0.08$\pm$ 0.01 & 0.08$\pm$ 0.02   & 0.0124 & 27.54 \\ \hline
 \end{tabular}
 \end{center}

 \newpage

\vspace*{10mm}
 \begin{figure}[h]
 \hspace*{-7mm}
 \includegraphics[scale=0.35]{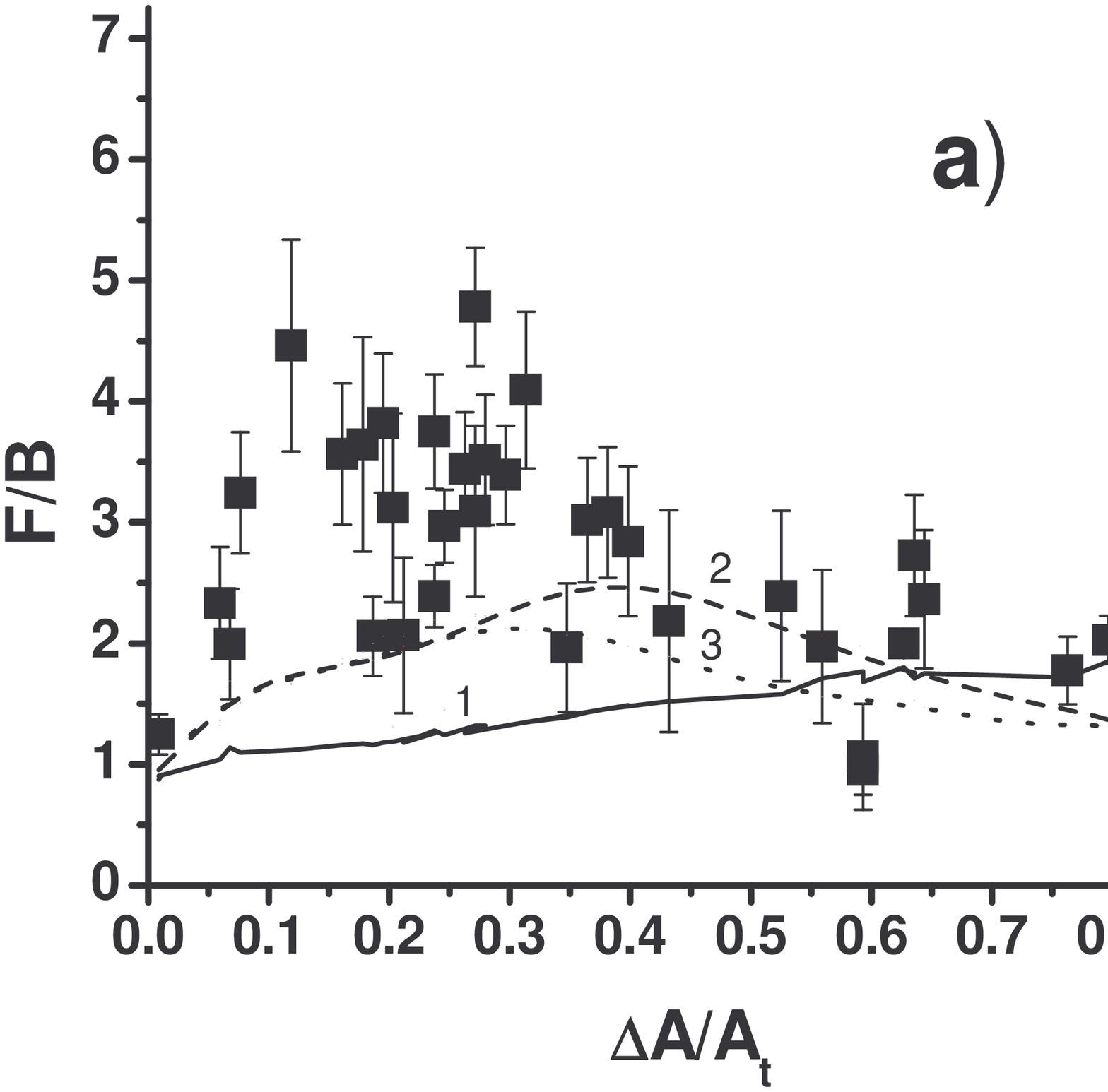}
 \hspace*{-15mm}
 \includegraphics[scale=0.35]{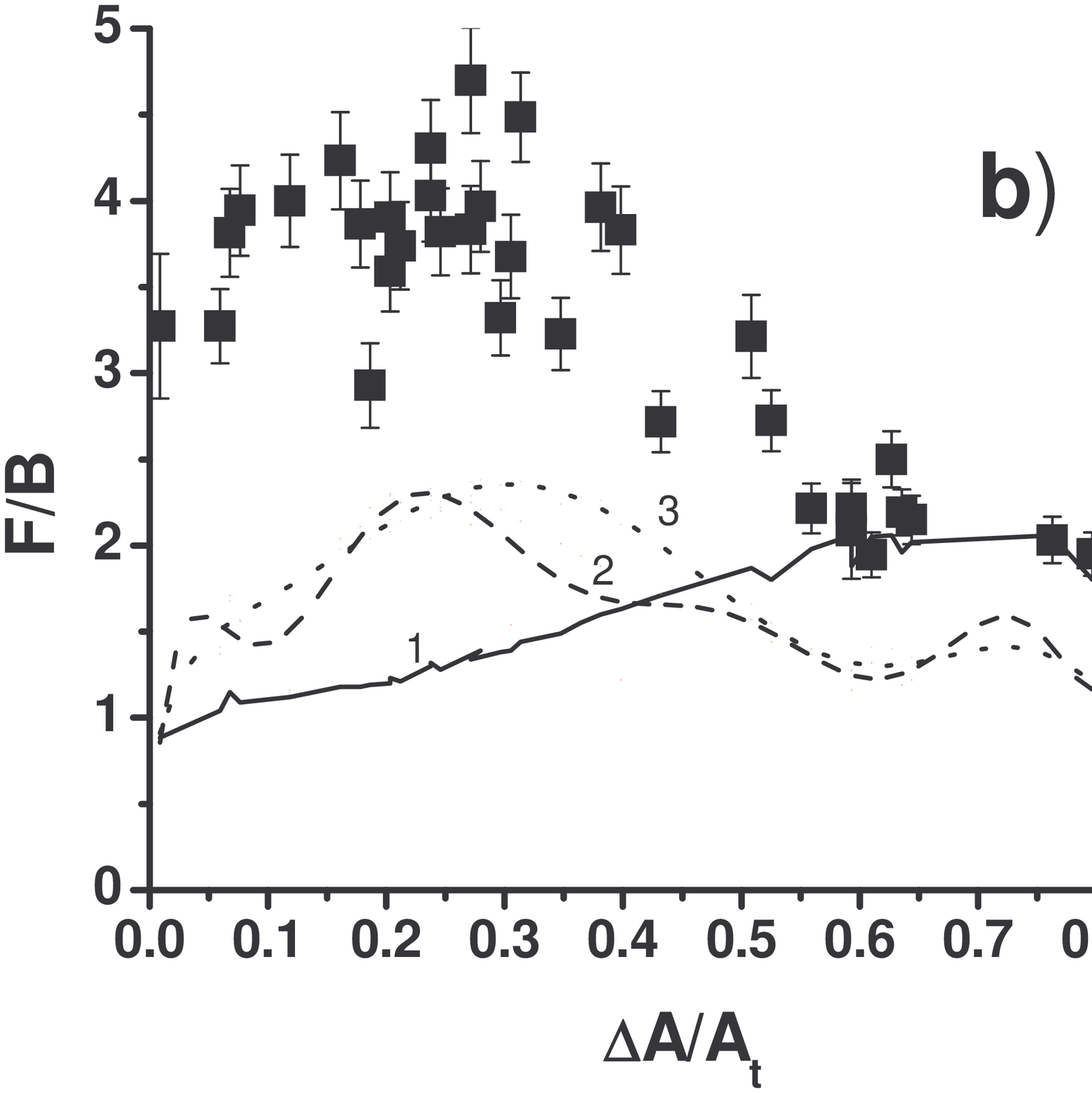}
 \vspace*{-5mm}
 \caption{
 $F/B$ versus the fractional mass losses $\Delta A/A_t$:
a) for deuteron-induced reactions b) for proton-induced
reactions. 
Solid lines (1) show calculations by LAQGSM03.01, dashed
lines (2) by LAQGSM03.S1, and dotted lines (3) 
by LAQGSM03.G1.
 }
 \end{figure}

 \newpage

\vspace*{10mm}
 \begin{figure}[h]
 \hspace*{-7mm}
 \includegraphics[scale=0.35]{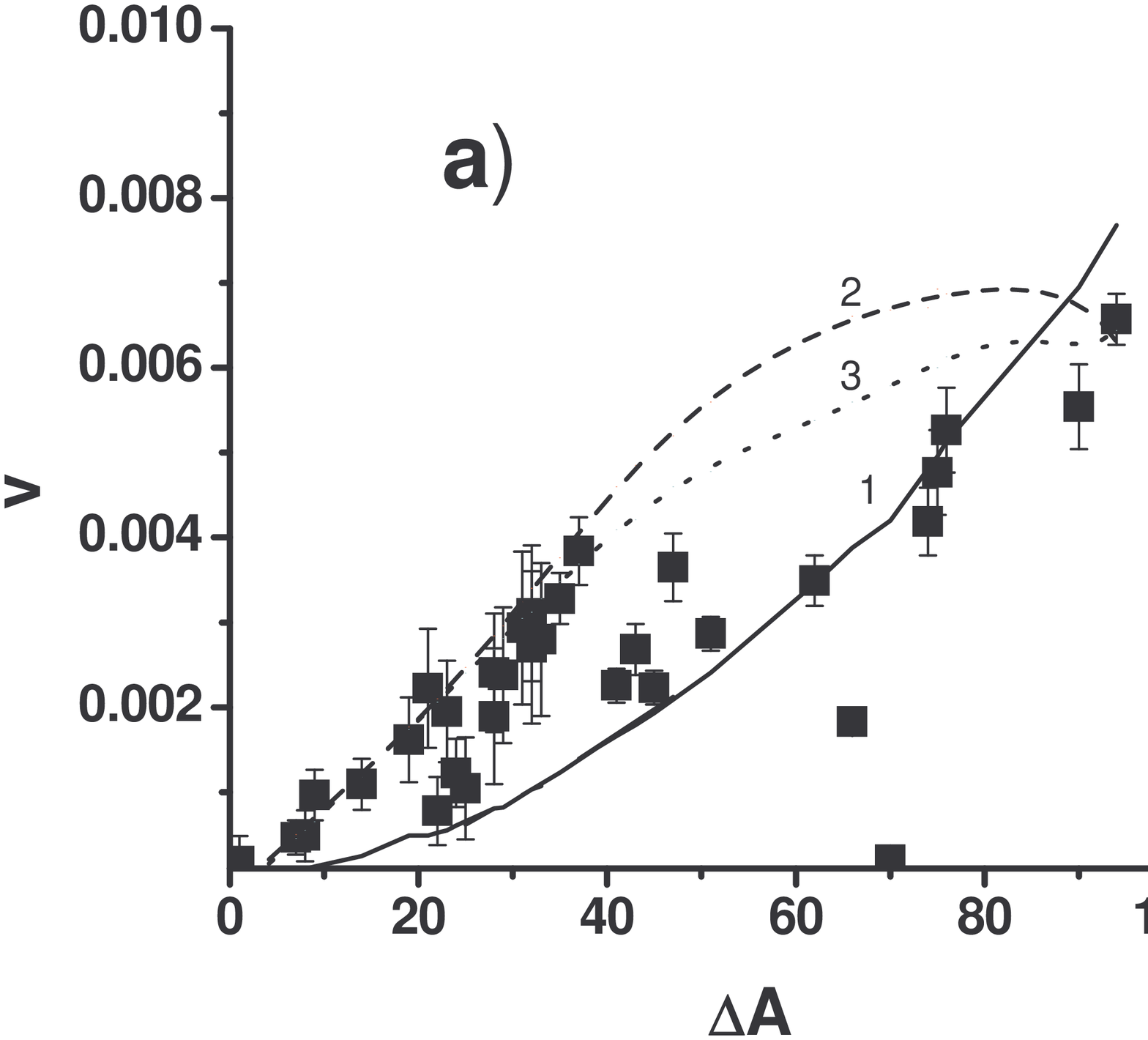}
 \hspace*{-15mm}
 \includegraphics[scale=0.35]{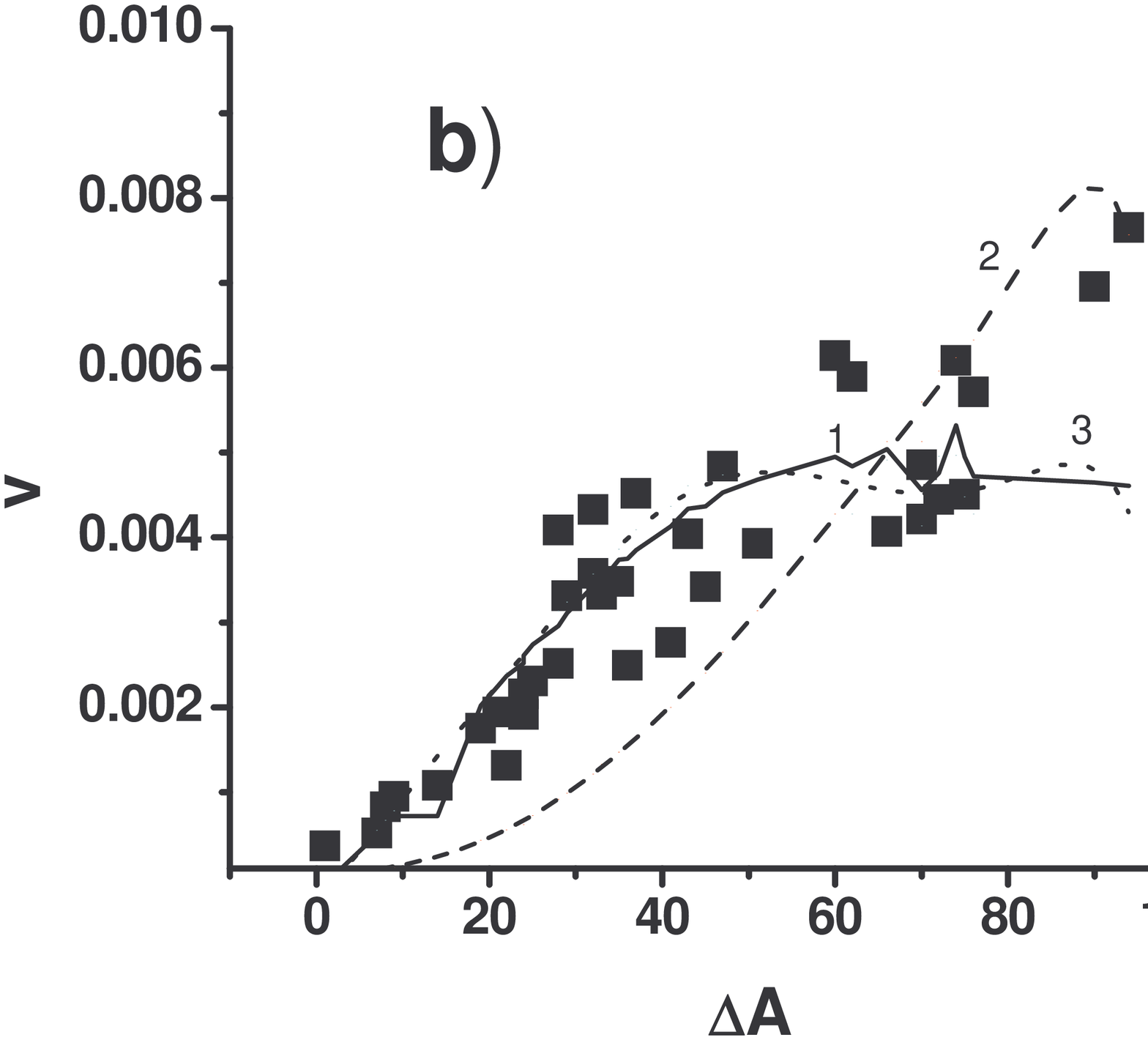}
 \vspace*{-5mm}
 \caption{
Dependence of forward velocity on the number of emitted
nucleons: a) for deuteron-induced reactions b) for proton-induced reactions. 
Solid lines (1) show calculations by LAQGSM03.01, dashed
lines (2) by LAQGSM03.S1, and dotted lines (3) 
by LAQGSM03.G1.
}
 \end{figure}

 \newpage

\vspace*{10mm}
 \begin{figure}[h]
 \hspace*{-7mm}
 \includegraphics[scale=0.35]{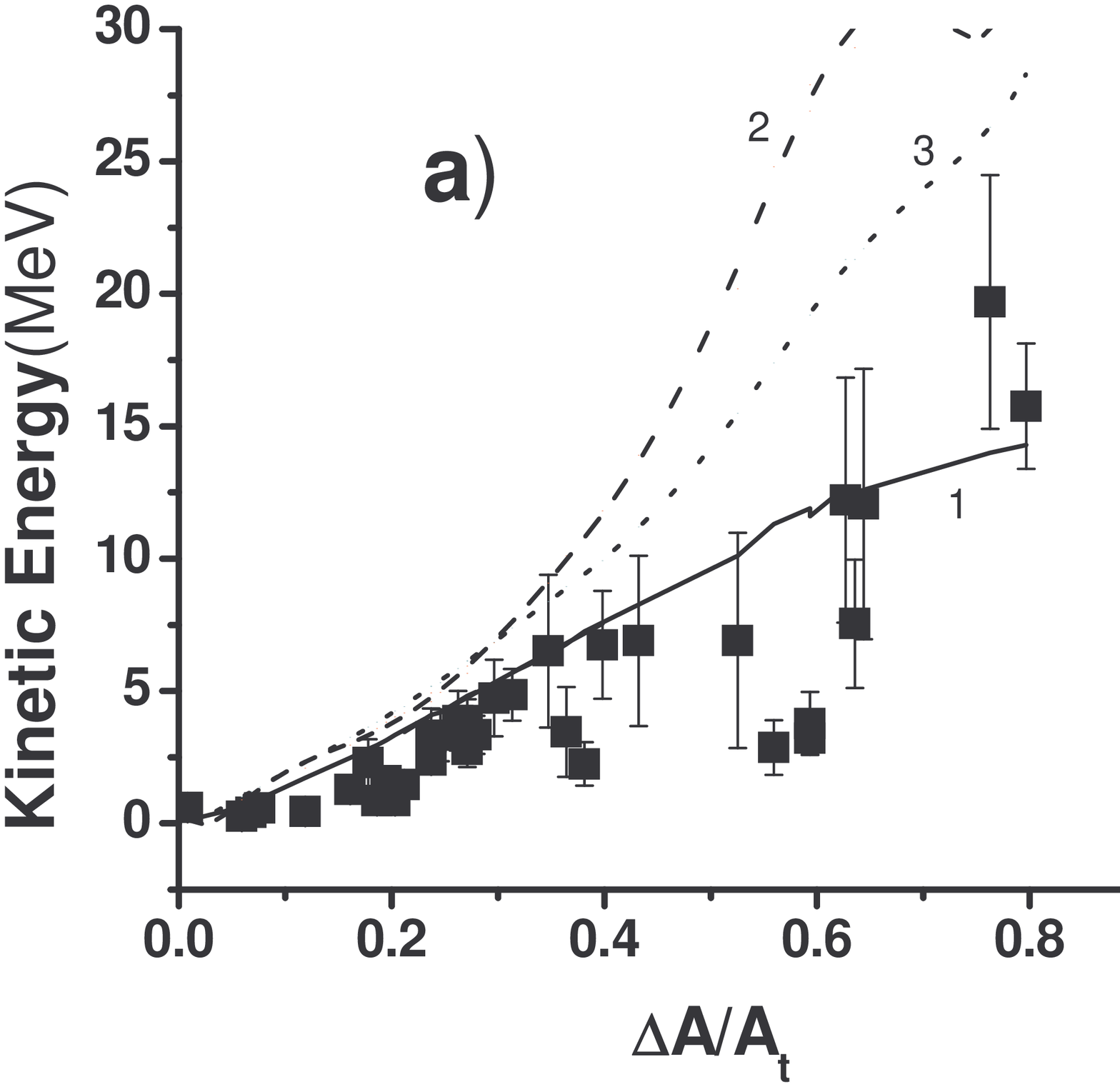}
 \hspace*{-15mm}
 \includegraphics[scale=0.35]{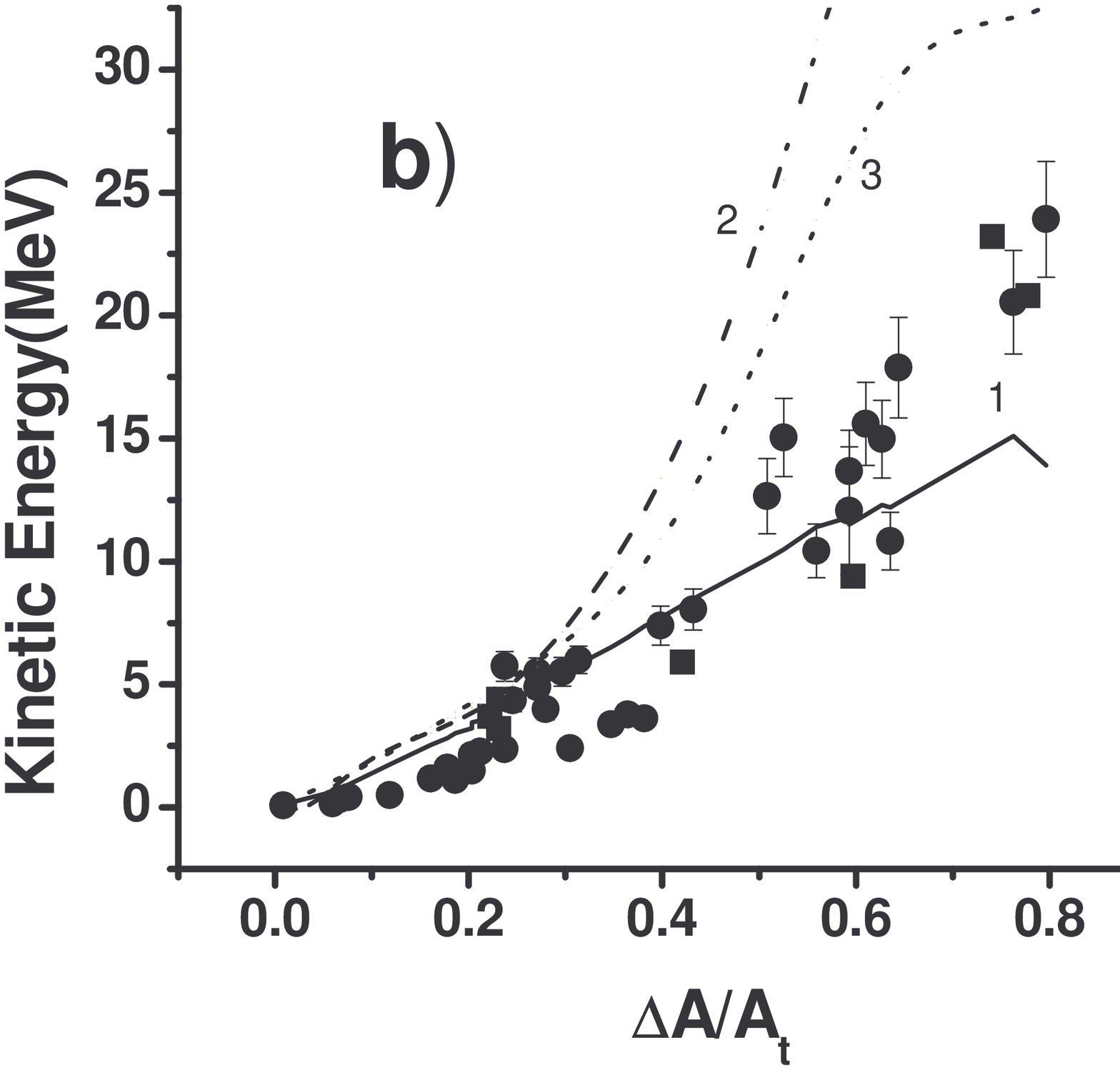}
 \vspace*{-5mm}
 \caption{
 Dependence of the kinetic energy of the product nuclei
on the fractional mass losses $\Delta A/A_t$: a) for 
deuteron-induced reactions 
b) for the proton-induced reactions ($\bullet$).
For comparison, 
$\blacksquare$ show experimental results for the Ag target
tabulated in Ref.\ \cite{win3}. 
Solid lines (1) show calculations by LAQGSM03.01,
dashed lines (2) by LAQGSM03.S1, and dotted lines (3)
by LAQGSM03.G1.
}
 \end{figure}

 \end{document}